\documentclass[12pt]{article}
\usepackage{epsfig, amsmath, amssymb}
\usepackage{graphicx,psfrag} 
\newcommand{\nn}{\nonumber}
\newcommand{\be}{\begin{equation}}
\newcommand{\ee}{\end{equation}}
\newcommand{\ben}{\begin{equation}}
\newcommand{\een}{\end{equation}}
\newcommand{\bea}{\begin{eqnarray}}
\newcommand{\eea}{\end{eqnarray}}
\newcommand{\bA}{\begin{array}}
\newcommand{\eA}{\end{array}}
\newcommand{\bc}{\begin{center}}
\newcommand{\ec}{\end{center}}
\newcommand{\al}{\alpha}

\newcommand{\ra}{\rightarrow}
\newcommand{\del}{\partial}

\newcommand{\ie}{{\it i.e.}}
\newcommand{\eg}{{\it e.g.}}

\def\BZ{{\mathbb Z}}

\newcommand{\lan}{\langle}
\newcommand{\ran}{\rangle}

\begin{document}


\begin{titlepage}

\bc

\hfill 

\vspace{20mm}


{\Huge On aspects of 2-dim dilaton gravity,
 \\ [2mm] 
dimensional reduction and holography}
\vspace{16mm}

{\large K.~Narayan} \\
\vspace{3mm}
{\small \it Chennai Mathematical Institute, \\}
{\small \it SIPCOT IT Park, Siruseri 603103, India.\\}

\ec
\vspace{35mm}

\begin{abstract}
  We discuss aspects of generic 2-dimensional dilaton gravity
  theories.  The 2-dim geometry is in general conformal to $AdS_2$ and
  has IR curvature singularities at zero temperature: this can be
  regulated by a black hole. The on-shell action is divergent: we
  discuss the holographic energy-momentum tensor by adding appropriate
  counterterms. For theories obtained by dimensional reduction of the
  gravitational sector of higher dimensional theories, for instance
  higher dimensional $AdS$ gravity as a concrete example, the
  2-dimensional description dovetails with the higher dimensional
  one. We also discuss more general theories containing an extra
  scalar field which now drives nontrivial dynamics. Finally we
  discuss aspects of the 2-dimensional cosmological singularities
  discussed in earlier work. These studies suggest that generic 2-dim
  dilaton gravity theories are somewhat distinct from JT gravity and
  theories ``near JT''.
\end{abstract}

\end{titlepage}

{\tiny   
\begin{tableofcontents}
\end{tableofcontents}
}

\vspace{-2mm}

\section{Introduction}

Dilaton gravity in 2-dimensions has been under active investigation in
recent years, in part following discussions of nearly $AdS_2$
holography \cite{Almheiri:2014cka}-\cite{Engelsoy:2016xyb} (reviewed
in \cite{Sarosi:2017ykf}-\cite{Trunin:2020vwy}) and more recently
those of Jackiw-Teitelboim (JT) gravity
\cite{Jackiw:1984je,Teitelboim:1983ux} being dual to ensembles
\cite{Saad:2019lba}, with further development in
\cite{Stanford:2019vob}-\cite{Alishahiha:2020jko}. It is well-known
that $AdS_2$ arises in the near horizon geometry of extremal black
holes and branes \cite{Sarosi:2017ykf}-\cite{Trunin:2020vwy}. However
2-dim dilaton gravity per se arises quite generically from the
2-dimensional subsector of higher dimensional gravity on spaces of the
form ${\cal M}_2\times X^d$ upon Kaluza-Klein compactification over
the compact space $X^d$.

It is interesting to ask how these generic
dilaton gravity theories compare with JT gravity, \ie\ if they are
``near JT''. Here we study certain aspects of generic 2-dim
dilaton-gravity theories of the form
\be\label{2dg-gen}
S= {1\over 16\pi G_2} \int d^2x\sqrt{-g}\, \big(\phi\mathcal{R}
-U(\phi) \big)\ ,
\ee
containing the 2-dim dilaton coupling to gravity and a general
dilaton potential. For JT gravity, the potential is\ $U=-2\phi$. From
the equations of motion,
\be\label{2dimseom-EMD0}
g_{\mu\nu}\nabla^2\phi-\nabla_{\mu}\nabla_{\nu}\phi
  +\frac{g_{\mu\nu}}{2}U = 0\ ,\qquad\quad
\mathcal{R}-\frac{\partial U}{\partial\phi} = 0\ .
\ee
For $U\sim \phi^n$ asymptotically, the curvature diverges when
the dilaton is vanishingly small:
\be
   {\cal R} \sim \phi^{n-1}\quad \xrightarrow{\ \phi\ra 0\ } \infty\ ,
   \qquad\quad n<1\ .
\ee
The 2-dim geometry is generically conformally $AdS_2$ and the conformal
factor leads to this (IR) singularity: this is in the zero
temperature theory. Another consequence of the dilaton equation of
motion is that the bulk on-shell action has a UV divergence where
$\phi$ grows large:
\be\label{onshellActionD}
S^{o.s.} = {1\over 16\pi G_2} \int d^2x\sqrt{-g}\,
\big(\phi\del_\phi U -U \big) \sim {1\over 16\pi G_2} \int d^2x\sqrt{-g}\,
(n-1) U\ .
\ee
By comparison, in JT gravity the metric is $AdS_2$ with constant
curvature, and the potential is linear so the on-shell action
vanishes.

In some sense, the theories (\ref{2dg-gen}) are reminiscent of certain
nonrelativistic theories: \eg\ hyperscaling violating (hvLif) theories
conformal to $AdS$ (or Lifshitz) exhibit curvature singularities in
the zero temperature metric\ (see
\cite{Taylor:2015glc,Hartnoll:2016apf} for reviews of various aspects
of nonrelativistic holography). These are best regarded as effective
gravity theories valid in certain intermediate regimes. For theories
with gauge/string realizations, a UV-complete description emerges in
the latter. For instance some of these can be obtained from the
supergravity description of nonconformal $Dp$-branes
\cite{Itzhaki:1998dd} upon dimensional reduction on the transverse
$S^{8-p}$ sphere \cite{Dong:2012se}. While the effective hvLif
theories here exhibit such singularities, the higher dimensional
description is well-behaved: the singularities arise from the
compactification.

So far our discussion of (\ref{2dg-gen}) is bottom-up. Looking
top-down, similar arguments apply at least when theories
(\ref{2dg-gen}) arise by dimensional reduction from well-behaved
theories in higher dimensions. A prototypical example in this regard
is the reduction of higher dimensional gravity with a cosmological
constant.  Restricting to just this gravity sector alone, and
focussing on a negative cosmological constant gives the action
\be\label{actionAdSD}
S_D = {1\over 16\pi G_D} \int d^Dx\sqrt{-g^{(D)}}
\left({\cal R}^{(D)}-2\Lambda\right) ,
\qquad\quad \Lambda=-{1\over 2}\,d_i(d_i+1) ,\qquad D=d_i+2\ .
\ee
We use reduction conventions of \cite{Kolekar:2018sba,Kolekar:2018chf,
  Bhattacharya:2020qil}\ (see also \cite{Strominger:1994tn,Nojiri:2000ja,
  Grumiller:2001ea,Grumiller:2002nm} for reviews).
Assuming translational \& rotational invariance in the $d_i$ boundary
spatial directions (for theories with holographic duals) leads to the
form\ ${\cal M}_2\times X^{d_i}$\ for the higher dimensional space.
For simplicity, we take $X^{d_i}$ to be a torus $T^{d_i}$ and compactify
with the reduction ansatz
\be\label{redux+Weyl}
ds^2_D = g^{(2)}_{\mu\nu} dx^\mu dx^\nu + \phi^{2\over d_i} d\sigma_{d_i}^2\ ;
\qquad\quad g_{\mu\nu}=\phi^{{d_i-1\over d_i}} g^{(2)}_{\mu\nu}\ .
\ee
The reduction and the Weyl transform $g^{(2)}\ra g$ above give the
2-dim dilaton gravity theory
\be\label{2dg-AdSD}
S= {1\over 16\pi G_2} \int d^2x\sqrt{-g}\, \big(\phi\mathcal{R}
-U(\phi) \big)\ ,\qquad\quad U=2\Lambda \phi^{1/d_i}\ .
\ee
See Appendix~\ref{sec:AppA} for some details. The 2-dim space here
contains time and the radial coordinate, while the $d_i=D-2$ spatial
coordinates define the ``transverse'' part of the bulk spacetime: in
theories with holographic duals, $d_i$ is the spatial dimension of the
boundary field theory. Note that $AdS_3$ also leads to $U=-2\phi$.
The dilaton defined as in (\ref{redux+Weyl}) is the
transverse area in the higher dim space\ $g_{ii}^{(D-2)/2}=\phi$.\
The higher dimensional background here is $AdS_D$ (Poincare) with\
$ds^2={1\over r^2}(-dt^2+dr^2)+{1\over r^2}dx_i^2$,\ of the general
form (\ref{redux+Weyl}).

The theory (\ref{2dg-AdSD}) is of the general form (\ref{2dg-gen}):
the 2-dim theory is conformally $AdS_2$ and the zero temperature
theory has a curvature singularity (sec.~\ref{sec:2dgGen}) and the
on-shell action is of the form (\ref{onshellActionD}) with $n={1\over
  d_i}$\,. Of course, operationally both of these are curable in
(\ref{2dg-AdSD}) and more generally in (\ref{2dg-gen})\,: the
curvature singularity is regulated by introducing a black hole horizon
(sec.~\ref{sec:BH}) and the on-shell action is regulated by adding
appropriate counterterms which render calculations of holographic
observables reasonable (sec.~\ref{sec:HolEM} and
sec.~\ref{sec:scalarProbes}).

These aspects are of course well-known in the extensive studies of
higher dimensional $AdS/CFT$
\cite{Maldacena:1997re}-\cite{Aharony:1999ti} over the years: here we
simply note that these 2-dim theories exhibit similar features.  In
particular, some aspects of these sorts of 2-dim dilaton gravity
theories have also been studied in \eg\ \cite{Taylor:2017dly}.  Among
other things, this describes the role in these theories of generalized
conformal structure \cite{Jevicki:1998ub} (a symmetry under scaling
transformations involving Weyl rescaling of the metric alongwith
appropriate scalings of additional scalars representing running
couplings), studied for nonconformal branes in
\cite{Kanitscheider:2008kd,Kanitscheider:2009as,Gouteraux:2011qh}
where various aspects of the holographic dictionary were described.
Recalling our earlier comments on nonconformal branes and
nonrelativistic theories, we see that the above features also apply in
some of the present discussions stemming from parallels with the
nontrivial dilaton profile which from (\ref{redux+Weyl}) is part of
the higher dimensional metric (with the associated higher dimensional
symmetries).

Thus overall, although 2-dimensional, (\ref{2dg-AdSD}) and more generally
(\ref{2dg-gen}) is apparently encoding higher dimensional gravity
intrinsically: this is quite different from the near extremal near
horizon $AdS_2\times X$ throats in extremal objects, where the
$X$-compactification in fact leads to an intrinsically 2-dim theory
with a clear separation of scales. To put this in perspective, we note
that the gravity subsector, taken stand-alone, is universal to all
string/M theories on $AdS_D\times X^{10/11-D}$. This is usually
understood to be UV complete only if the entire higher dimensional
(string/M) upstairs theory is included. From this point of view, it
should not be surprising that these 2-dim theories exhibit the
divergences above: they are perhaps best regarded as UV-incomplete low
energy effective theories universal to all UV completions $AdS_D\times
X$ upstairs, so they are thermodynamic, akin to an ensemble.  These
features appear generic to 2-dim dilaton gravity theories of the form
(\ref{2dg-gen}), with general (nonlinear) potentials. They also arise
with additional matter, \eg\ extra scalars that drive dynamics
(sec.~\ref{sec:Psi}). These studies suggest that generic 2-dim dilaton
gravity theories are somewhat distinct from JT gravity and theories
``near JT'' which appear special. In what follows we will describe
this in greater detail, with a Discussion in sec.~\ref{sec:Disc}.

\section{Aspects of generic 2-dim dilaton gravity}\label{sec:2dgGen}

We now describe in more detail aspects of the 2-dim theory 
(\ref{2dg-gen}). Using conformal gauge and combining the various
Einstein equation components, the equations of motion
(\ref{2dimseom-EMD0}) become
\bea\label{2dimseom-EMD1}
&& \qquad\qquad ds^2 = g_{\mu\nu}dx^\mu dx^\nu = e^{f(r)} (-dt^2+dr^2)\ ,
\nn\\ [1mm]
&& - \del_r^2\phi + \del_rf\,\del_r\phi = 0\,,\qquad
\del_r^2\phi + e^f U = 0\,,\qquad
- \del_r^2f- e^f\del_\phi U = 0\,.
\eea
It is adequate to exclude time dependence here: later we will discuss
nontrivial time dynamics as well in the presence of an extra
scalar. Solutions to these can be expressed in terms of a general
``pre-potential $\int d\phi' U(\phi')$\ \cite{Cavaglia:1998xj} (see
also \cite{Maldacena:2019cbz}, Appendix E).  In cases where $U(\phi)$
is a $\phi$-monomial, solutions can be found by employing a power-law
scaling ansatz for $\phi, e^f$, and then solving the algebraic
equations that result from (\ref{2dimseom-EMD1}): this was found to be
useful in \cite{Bhattacharya:2020qil}. For the theory (\ref{2dg-gen}),
a monomial potential gives from (\ref{2dimseom-EMD1})
\bea\label{X^2e^f-gen}
U=A\phi^n ;\qquad\qquad \phi=r^m ,\ e^f=r^b \qquad \Rightarrow
\qquad\qquad\qquad\qquad\qquad\qquad  \nn\\ [1mm]
\big(-m(m-1) + bm\big) r^{m-2} = 0 ,\ \ m(m-1)r^{m-2} + A r^{b+mn} = 0 ,
\ \ b r^{-2} - An r^{b+m(n-1)} = 0 , \nn\\ [1mm]
\Rightarrow\qquad  m(b-m+1)=0\,,\quad  b+m(n-1)+2 = 0\,,\quad
m(m-1)=-A\,,\quad b=An\,.\quad
\eea
We require the $r$-exponents to match for a nontrivial solution valid
for all $r$: this gives \eg\ $m-2=b+mn$, and the remaining equations
then follow. (The lengthscale we have suppressed can be reinstated
when required)\ With $m\neq 0$, we obtain $m=b+1$ and thereby
\be\label{mb-gen}
m=-{1\over n}\,,\qquad b=-{1\over n}-1\,,\qquad {\rm and}\qquad
A = -{1\over n} \Big({1\over n}+1\Big)\ ,
\ee
as the unique solution, which can be checked to satisfy all the
equations in (\ref{X^2e^f-gen}).
The last equation here is a condition that the parameters $A,n$, in
the potential $U$ must satisfy for consistency of this solution to
the theory.   
Thus the 2-dim background solution to (\ref{2dg-gen}) is
\be\label{2dsoln-gen}
\phi = {1\over r^{1/n}}\,, \qquad\quad
ds^2 = {1\over r^{1+{1\over n}}} (-dt^2+dr^2)
= {1\over r^{{1\over n}-1}}\, ds^2_{AdS_2}\,.
\ee
The curvature (\ref{2dimseom-EMD0}) is
\be
{\cal R} = An\phi^{n-1} = -\Big(1+{1\over n}\Big)\,\, r^{{1\over n}-1}\ ,
\ee
giving an IR singularity in the deep interior at $r\ra\infty$ for $n<1$
and at $r\ra 0$ if $n>1$. JT gravity is the case $n=1$, the space
being $AdS_2$, the metric being $ds^2={1\over r^2}(-dt^2+dr^2)$ 
with constant curvature ${\cal R}=-2$.

Thus for $n\neq 1$, the 2-dim geometry is conformal to $AdS_2$:
for $n<1$ the conformal factor becomes larger towards the boundary
$r\ra 0$, so the space there ``flattens''. As $\phi$ grows to large
field values, the curvature ${\cal R}\ra 0$.
The conformally $AdS_2$ geometry (\ref{2dsoln-gen}) is akin to
a nonconformal $Dp$-brane geometry \cite{Itzhaki:1998dd}: the
corresponding Euclidean space is
\be\label{confAdS2}
ds^2_E
\sim d\rho^2 + \rho^{2(1+n)\over 1-n} dt_E^2\ ,
\qquad \rho \sim r^{{1-1/n\over 2}}\,;
\qquad {\cal R}\sim -{1\over\rho^2}\ .
\ee
In general a conformally $AdS_2$ metric $ds^2={e^{F(r)}\over
  r^2}(dt_E^2+dr^2)$ has curvature ${\cal R}=-e^{-F}(2+r^2F'')$ with a
singularity as $e^F\ra 0$\ (as $r\ra\infty$) unless $F=2\log r$ (flat
space) or $F=0$ ($AdS_2$).\ 
More general dilaton potentials can be used to find classical solutions
to the equations (\ref{2dimseom-EMD0}): in general they will lead to
conformally $AdS_2$ spaces with singularities.

\subsection{Regulating with a black hole}\label{sec:BH}

There are parallels with the IR curvature singularity described above
for $n<1$ and similar singularities in hyperscaling violating theories:
on the left below is the zero temperature metric
\be\label{hvLif}
ds^2=r^{2\theta/d_i} \Big({-dt^2+dr^2+dx_i^2\over r^2}\Big)
\quad\longrightarrow\quad
ds^2={r^{2\theta/d_i}\over r^2} \Big(-f(r)dt^2+dx_i^2+{dr^2\over f(r)}\Big)\ .
\ee
Here the exponent $\theta<0$ and the conformal factor leads to a
curvature singularity ${\cal R}\sim r^{-2\theta/d_i}$ as
$r\ra\infty$. More general nonrelativistic theories include a Lifshitz
exponent as well in the form\ $-{dt^2\over r^{2z}}+\ldots$ above and
$\theta$ can be positive as well. These also have singularities with
either curvatures or tidal forces diverging: see
\cite{Taylor:2015glc,Hartnoll:2016apf} which review various aspects of
nonrelativistic holography.  In general these are best regarded as
effective theories valid in some intermediate regimes: beyond these
some further completion is required. In fact nonconformal $Dp$-branes
upon reduction on the transverse $S^{8-p}$ sphere give (\ref{hvLif})
with $\theta=p-{9-p\over 5-p}$\ and $d_i=p$\ \cite{Dong:2012se}: the
higher dimensional description is of course well known in string
theory, with rich phase diagrams \cite{Itzhaki:1998dd}. Similar
features arise in the string realizations in \cite{Narayan:2012hk}.

At the low energy effective level, the metric on the left above can
be de-singularized by introducing a black hole/brane: the metric on
the right in (\ref{hvLif}) is at finite temperature with a blackening
factor $f(r)\sim 1-{r^\#\over r_0^\#}$ that vanishes at a horizon
$f(r_0)=0$. Now the space is smooth at the Hawking periodicity,
where the Euclidean geometry has no conical singularity.

The behaviour of the conformally $AdS_2$ space (\ref{confAdS2}) is
very similar: the zero temperature theory is singular at the IR horizon
$r\ra\infty$ due to the conformal factor. As above, we can regulate this
via a black hole horizon,
\be\label{AdSDbb1}
ds^2 = e^{f(r)}(-dt^2+dr^2)\ \ \longrightarrow\ \
e^{f(r)} \Big( -H(r) dt^2 + {dr^2\over H(r)} \Big)\,;\quad
{\rm with}\ \ H\xrightarrow{r\ra 0} 1,\ \ H\xrightarrow{r\ra r_0} 0\ .
\ee
The Einstein equations (\ref{2dimseom-EMD0}) now give\footnote{
We have\ $\Gamma^t_{tr} = {f'\over 2}+{H'\over 2H}\,,\ \
\Gamma^r_{rr} = {f'\over 2}-{H'\over 2H}\,,\ \
\Gamma^r_{tt} = {H\over 2} (Hf' + H')\,,\quad 
{\cal R} = -e^{-f} \big(Hf'' + H'f' + H''\big)$. }
\be
-\del_r^2\phi+f'\del_r\phi=0 ,\qquad \del_r(H\del_r\phi)+e^fU=0\ ,
\ee
The first equation implies\ $\del_r\phi\propto e^f$.\ Now since we are
thinking of the blackening factor $H(r)$ as a finite temperature
regulator to the original space, these equations must hold
simultaneously alongwith (\ref{2dimseom-EMD1}): after imposing the
$H$ boundary conditions in (\ref{AdSDbb1}), this gives
\be\label{AdSDbb2}
\del_r(H\del_r\phi)=-e^fU = \del_r^2\phi\ \ \ra\ \
\del_r\big((H-1)\del_r\phi\big)=0\ \ \ra\ \
H=1-e^{-f+f_0}\ .
\ee
Near the horizon $r\sim r_0$, the Euclidean space becomes\ ($H'_0=f'_0<0$)
\be
ds^2\xrightarrow{r\ra r_0}\,
e^{f_0}\Big(H'|_0(r-r_0)dt_E^2+{dr^2\over H'|_0(r-r_0)}\Big)
\ra\ {f_0'^2\over 4}\rho^2dt_E^2+d\rho^2\,;\quad\ \&\ \ \phi\ra\phi_h\ ,
\ee
which has no conical singularity if\ ${t_E|f'_0|\over 2}$ has
periodicity $2\pi$ giving the Hawking temperature
$T={1\over\beta}={|f'_0|\over 4\pi}$\,. This regulates the IR singularity.
The Lorentzian geometry as usual has an eternal extension with two
asymptotic regions connected by an Einstein-Rosen bridge (wormhole),\
and a spacelike curvature singularity as $r\ra\infty$ cloaked by
the horizon as usual, with ${\cal R}\sim r^{1/n-1}$.

To desingularize (\ref{2dsoln-gen}), we could also add charge and
tune to extremality but this requires adding a gauge field, whereas
the regulator above is contained within the theory (\ref{2dg-gen}).

\subsubsection{$AdS_D$ gravity reduction}

The above general discussions are readily seen to hold in the
theory (\ref{2dg-AdSD}), with
\be
n={1\over d_i}\ ,\qquad\qquad A=2\Lambda=-d_i(d_i+1)\ ,
\ee
which arises from the reduction of $AdS_D$ gravity.
Then the solution (\ref{2dsoln-gen}) becomes
\be\label{AdSD-2dmetDil}
\phi={1\over r^{d_i}}\,,\qquad
ds^2={1\over r^{d_i+1}} (-dt^2+dr^2)={1\over r^{d_i-1}} ds^2_{AdS_2}\ .
\ee
\ie\ $b=-d_i-1,\ m=-d_i$. From (\ref{redux+Weyl}), we see this to be
the reduction of $AdS_D$ Poincare, \ie\ 
\be
ds_D^2={1\over r^2}(-dt^2+dr^2)+{1\over r^2}dx_i^2\ ,
\ee
suppressing the $AdS$ scale.
The compactification from higher dimensions gives the conformal factor
which leads to the IR curvature singularity in ${\cal R}$\
(\ref{confAdS2}) at the horizon $\rho=0$ (\ie\ $r\ra\infty$). Now the
black hole regulator (\ref{AdSDbb1}), (\ref{AdSDbb2}), in this case
(\ref{AdSD-2dmetDil}), gives the 2-dim space
\be\label{AdSDbb3}
\phi={1\over r^{d_i}}\,;\quad
ds^2 = {1\over r^{d_i+1}} \Big( -H(r) dt^2 + {dr^2\over H(r)} \Big) ,\quad
H(r) = 1-{r^{d_i+1}\over r_0^{d_i+1}}\,;\qquad T={d_i+1\over 4\pi r_0}\ .
\ee
This is essentially the dimensional reduction of the $AdS_D$ black
brane.

This sort of IR horizon singularity is generic for such compactifications:
for instance the reduction of global $AdS_D$\ (see
Appendix~\ref{sec:AppA}) gives
\be
ds^2=-(1+r^2)dt^2+{dr^2\over 1+r^2}+r^2d\Omega_i^2 \ \ra\
\phi=r^{1/d_i},\ \ e^f=r^{d_i-1} \Big(-(1+r^2)dt^2+{dr^2\over 1+r^2}\Big) ,
\ee
with a curvature singularity in the deep interior $r\ra 0$\
(here $r\ra\infty$ is the boundary).

Aspects of the torus reduction in (\ref{AdSD-2dmetDil}) above were
also discussed in \cite{Taylor:2017dly} in light of generalized
conformal structure: more general reductions can also be carried out
where each of the torus directions is associated with a distinct warp
factor (see also \cite{Gouteraux:2011qh}).

\section{The on-shell action and the holographic stress tensor}
\label{sec:HolEM}

As mentioned earlier, the on-shell action (\ref{onshellActionD}) for
(\ref{2dg-gen}) has a UV divergence for $n\neq 1$. For
(\ref{2dg-AdSD}), this in fact just descends from the known $AdS_D$
on-shell action. We will now discuss the holographic energy-momentum
tensor focussing on (\ref{2dg-AdSD}) which allows direct comparison
with the higher dimensional description: many of the expressions are
valid for (\ref{2dg-gen}) as well, with $d_i\ra{1\over n}$ as will be
obvious from context. Our discussion is of course motivated by the
well-known formulations of holographic renormalization
\cite{Balasubramanian:1999re,Myers:1999psa,de
  Haro:2000xn,Skenderis:2002wp} in higher dimensional theories: see
also \cite{Cvetic:2016eiv} in certain 2-dim theories and
\cite{Kanitscheider:2008kd} in nonconformal branes.

For an $AdS$-like metric, with timelike boundary $\del{\cal M}$ at $r=0$
and outward pointing normal $n_r<0$ with $\mu,\nu\in\del{\cal M}$, the
extrinsic curvature at $\del{\cal M}$ is given by
\bea\label{nK-2d}
&& ds^2_D = N^2dr^2+\gamma_{\mu\nu}dx^\mu dx^\nu\,,\quad\ n=-Ndr,\quad\
K_{\mu\nu}=-{1\over 2}(\nabla_\mu n_\nu+\nabla_\nu n_\mu)=\Gamma_{\mu\nu}^rn_r\,;
\nn\\
&& ds^2 = e^f(-dt^2+dr^2)\ \Rightarrow\ N^2=-\gamma_{tt}=e^f ;\
K_{tt}={1\over 2N}\del_r\gamma_{tt}\,,\ \
K=\gamma^{tt}K_{tt}={f'\over 2}e^{-f/2}\,.\qquad
\eea
The renormalized action, comprising the bulk and the Gibbons-Hawking
boundary terms, is
\be\label{Sren}
S_{ren} = {1\over 16\pi G_2} \left[
  \int d^2x\sqrt{-g}\, \big(\phi{\cal R}-U(\phi)\big)
  - 2 \int dt \sqrt{-\gamma}\, \phi K
  - 2 \int dt \sqrt{-\gamma}\,\phi^{{d_i+1\over 2d_i}}\,d_i \right] ,
\ee
as well as a holographic counterterm $S_{ct}$\ (see also
\cite{Cvetic:2016eiv}), which cancels the divergences as
\be\label{SD+ct-div}
S_{ren}^{div}\ \propto\ {1\over 16\pi G_2} \left[
  {(d_i-1)\over\epsilon^{d_i+1}} + {(d_i+1)\over\epsilon^{d_i+1}}
- {2d_i\over\epsilon^{d_i+1}} \right] \ra\ 0\ ,
\ee
thus regulating the gravitational action. Using the boundary
metric from the ansatz (\ref{redux+Weyl}), the counterterm can be seen
to be just the reduction of the familiar $AdS_D$ counterterm\
$\int \sqrt{h^{(d_i+1)}}\ra \int dt\sqrt{\gamma^{(2)}} g_{ii}^{d_i/2}
= \int dt \sqrt{e^f\,\phi^{-(d_i-1)/d_i}}\, \phi$, which gives the
above.
The variation of the regulated action (\ref{Sren}) with the metric
can be written as \cite{KolekarThesis,Grumiller:2017qao}
\bea\label{Sren-varn}
\delta S_{ren} &=& {1\over 16\pi G_2} \int d^2x\sqrt{-g} 
       \left( g_{\mu\nu}\nabla^2\phi-\nabla_{\mu}\nabla_{\nu}\phi
  +\frac{g_{\mu\nu}}{2}U \right) \delta g^{\mu\nu} \nn\\
  && \qquad -\, {1\over 16\pi G_2} \int dt\sqrt{-\gamma}\,
  \gamma_{\mu\nu}\,n^\rho\nabla_\rho\phi\,\delta\gamma^{\mu\nu}\,
       +\, {\delta S_{ct}\over\delta \gamma^{\mu\nu}} \delta \gamma^{\mu\nu}\ .
\eea
The bulk variation term (obtained after a cancellation of second
derivative terms between ${\cal R}$ and $K$) leads to the bulk
equations of motion which vanish on-shell.
The boundary terms above satisfy $n_\rho\delta\gamma^{\rho\sigma}=0$.
This is consistent with \cite{Almheiri:2014cka}.
Noting $n_r=-N$ from (\ref{nK-2d}), the variation of the boundary
metric then gives the holographic energy-momentum tensor as (after
rescaling by the $\epsilon$-factors in $\sqrt{-\gamma}\gamma^{tt}$
to match the boundary value $\sqrt{-{\hat\gamma}}{\hat\gamma}^{tt}$)
\be\label{Tttren}
T^{ren}_{tt} = -{2\over\sqrt{-{\hat\gamma}}}{\delta S_{ren}
  \over\delta{\hat\gamma^{tt}}} = {\epsilon^{{d_i+1\over 2}}\over 8\pi G_2} \left(
-\sqrt{g^{rr}} \del_r\phi - d_i\phi^{{d_i+1\over 2d_i}} \right) \gamma_{tt}\ .
\ee
This is regarded as an expansion in $\epsilon$ where the divergent
term is cancelled by the counterterms leaving behind possible finite
pieces with higher order pieces vanishing when the cutoff is removed
as $\epsilon\ra 0$.\ For the $AdS_D$ reduction (\ref{AdSD-2dmetDil}),
there are no subleading terms and the counterterms have been
engineered to cancel the divergences, so $T^{ren}_{tt}$ vanishes:
\be\label{Tttren-AdSD}
T^{ren}_{tt} = {\epsilon^{{d_i+1\over 2}}\over 8\pi G_2}
\left(-\epsilon^{{d_i+1\over 2}}\, {(-d_i)\over\epsilon^{d_i+1}} -
     {d_i\over\epsilon^{{d_i+1\over 2}}} \right) {-1\over\epsilon^{d_i+1}} = 0\ .
\ee
The vanishing stress tensor indicates that the background metric and
dilaton define the ``vacuum'' of the theory: the stress tensor
(\ref{Tttren}) essentially encodes excitations about this vacuum.
For the black hole in this theory (\ref{AdSDbb3}),
(\ref{AdSDbb1}), (\ref{AdSDbb2}),
which is the reduction of the $AdS_D$ black brane, the subleading
blackening term in $g^{rr}$ gives a nonzero energy-momentum 
\be\label{Tttren-AdSDbb}
T^{ren}_{tt} = \lim_{\epsilon\ra 0}\ {\epsilon^{{d_i+1\over 2}}\over 8\pi G_2}
\left(-\epsilon^{{d_i+1\over 2}}\, {(-d_i)\over\epsilon^{d_i+1}}
\sqrt{1-{\epsilon^{d_i+1}\over r_0^{d_1+1}}}
- {d_i\over\epsilon^{{d_i+1\over 2}}} \right) {-1\over\epsilon^{d_i+1}}\,
=\, {d_i\over 16\pi G_2\,r_0^{d_i+1}}\ .
\ee
We will now discuss the holographic energy-momentum tensor in the
higher dimensional theory \cite{Balasubramanian:1999re,
  Myers:1999psa,de Haro:2000xn,Skenderis:2002wp}, written in the
variables of the 2-dim
theory (\ref{redux+Weyl}) obtained after reduction. So consider
\bea\label{Ddim-metK}
&& ds^2_D = G_{rr}dr^2 + g^{(2)}_{tt} dt^2 + \phi^{2\over d_i} dx_i^2
\equiv G_{rr}dr^2 + h_{\mu\nu}dx^\mu dx^\nu\ ,\nn\\
&& K_{\mu\nu} = \Gamma_{\mu\nu}^rn_r={1\over 2}\sqrt{G^{rr}} \del_rh_{\mu\nu}\,,
\qquad 
K = {1\over 2}\sqrt{G^{rr}}\Big(h^{tt}\del_rh_{tt}+2{\del_r\phi\over\phi}\Big) ,
\eea
simplifying the $ii$-part in $K$. Then the boundary stress tensor
with the usual counterterm $\sim\int d_i\sqrt{-h}$\ (rescaled by the
$\epsilon$-factors in $\sqrt{-h}h^{\mu\nu}$ to match the boundary
value $\sqrt{-{\hat h}}{\hat h}^{\mu\nu}$) is
\be\label{Ttt-D}
T^{(D)}_{\mu\nu} = - {2\over\sqrt{-{\hat h}}}{\delta S_D\over\delta
  {\hat h}^{\mu\nu}}
= {\epsilon^{1-d_i}\over 8\pi G_D} (K_{\mu\nu}-Kh_{\mu\nu}-d_ih_{\mu\nu})
\ \ \ra\ \
T^{(D)}_{tt} = {\epsilon^{1-d_i}\over 8\pi G_D} \Big(
- \sqrt{G^{rr}}\,{\del_r\phi\over\phi} - d_i \Big) h_{tt}
\ee
Using (\ref{Ddim-metK}) and going to the variables in the
Weyl-transformed frame (\ref{redux+Weyl}) gives
\be
T^{(D)}_{tt} = {\epsilon^{1-d_i}\over 8\pi G_D} \left(
- \sqrt{g^{rr}}\,\del_r\phi - d_i\phi^{{d_i+1\over 2d_i}} \right)
\phi^{-{d_i+1\over 2d_i}}\, \gamma_{tt}\,\phi^{-{d_i-1\over d_i}} 
\ee
noting $\sqrt{G^{rr}}=\sqrt{g^{rr}\phi^{{d_i-1\over d_i}}}$ and
$\phi|_B={1\over\epsilon^{d_i}}$\,:\ this matches (\ref{Tttren}).\
Note that reduction of the higher dimensional theory directly gives
the holographic stress tensor in the 2-dim theory:
using the 2-dim variables (\ref{redux+Weyl}), we have
\bea
\delta S^{grav} &\sim& -{1\over 16\pi G_D} \int d^{d_i+1}x \sqrt{-h}\,
(K_{\mu\nu}-Kh_{\mu\nu}) \delta h^{\mu\nu} \nn\\
&=& -{V_{d_i}\over 16\pi G_D} \int dt  \sqrt{-\gamma^{(2)}}\, \phi
\left(K_{tt}-Kh^{(2)}_{tt}\right) \delta h_{(2)}^{tt} + \ldots\nn\\
&=& {1\over 16\pi G_2} \int dt \sqrt{-\gamma}\, \sqrt{g^{rr}} \del_r\phi\,
\gamma_{tt}\ \delta\gamma^{tt} + \ldots
\eea
noting that the $tt$-terms in $K$ cancel with $K_{tt}$ and the Weyl
factors cancel between $\sqrt{G^{rr}}$
and $\sqrt{-\gamma^{(2)}}$, as well as between $h^{(2)}_{tt},\
\delta h_{(2)}^{tt}$.\ 
For the $AdS_D$ black brane, (\ref{Ttt-D}) gives
\be
ds^2 = \phi^{-{d_i-1\over d_i}} e^f \Big(-Hdt^2+{dr^2\over H}\Big)
+ \phi^{{2\over d_i}} dx_i^2\quad\ra\quad
T^{(D)}_{tt}={d_i\over 16\pi G_D r_0^{d_i+1}}\ ,
\ee
matching the 2-dim density (\ref{Tttren-AdSDbb}) upon including
the compactification volume\ ${V_{d_i}\over G_D}\ra {1\over G_2}$\,.

It is useful to note that the 2-dim energy-momentum tensor calculation
is in the Weyl frame where the dilaton kinetic term has been absorbed
away: this matches with the higher dimensional calculation above.  In
a different 2-dim Weyl frame with the dilaton kinetic term present,
further holographic counterterms need to be added to cancel
divergences, which then will lead to the same conclusions.

\subsection{Low-lying (``soft'') modes}\label{sec:soft}

It is interesting to ask if there is any analog of the $nAdS_2$
Schwarzian action for low-lying fluctuations (``soft modes'') for
the Euclidean theory (\ref{2dg-AdSD}), \ie\ conformally $AdS_2$.
To study this, we mimic \cite{Maldacena:2016upp} and imagine small
boundary wiggles with boundary time $u$ and tangent $t^\mu(u)$ and
normal $n^\mu(u)$: this gives the general expression for any
conformally $AdS_2$ metric\footnote{For the Euclidean space
  $ds^2=e^f(d\tau^2+dr^2)$, we have\
  $\Gamma^\tau_{\tau r} = \Gamma^r_{rr} = -\Gamma^r_{\tau\tau}
  = {f'\over 2} = -{(d_i+1)/2\over r}$\,,\ and
$t^\mu=(\tau',r'),\ \ n^\mu={e^{-f/2}\over\sqrt{\tau'^2+r'^2}}(-r',\tau')$\,,\
with\ $g_{\mu\nu}n^\mu n^\nu=1$\ and\ $\del_u=\tau'\del_\tau+r'\del_r$.}
\be\label{Ksoft1}
K = -{t^\mu t^\nu \nabla_\nu n_\nu\over t^\mu t_\mu}
= -{e^{-f/2}\over (\tau'^2+r'^2)^{3/2}}
\left( r'\tau''-\tau'r''+{f'\over 2}\tau'(\tau'^2+r'^2) \right) .
\ee
The natural boundary conditions on the (Euclidean) conformally $AdS_2$
metric (\ref{AdSD-2dmetDil}) are
\be\label{confAdS2-bc}
{{\tau'}^2+{r'}^2\over r^{d_i+1}}={1\over\epsilon^{d_i+1}}\qquad \ra\qquad
r\sim \epsilon(\tau')^{{2\over d_i+1}}\ ,
\ee
to leading order. This gives\footnote{I thank Kaberi Goswami and
  Hitesh Saini for correcting an error here.}
\bea\label{Ksoft2}
&& K = \epsilon^{{d_i-1\over 2}} (\tau')^{{d_i-1\over d_i+1}} \left[
   {d_i+1\over 2}\ +\
   \epsilon^2\ \Big({2\over d_i+1}\Big) (\tau')^{{-4d_i\over d_i+1}} \left(
 \tau' \tau''' - {5d_i+1\over 2(d_i+1)} (\tau'')^2 \right)\right.\qquad \nn\\
         [1mm]
&& \left. \qquad\qquad\qquad\qquad\ 
     +\,\ \epsilon^4\, \Big({12\over (d_i+1)^3}\Big)\,
      (\tau')^{{-8d_i\over d_i+1}}\, (\tau'')^2
      \left( \tau' \tau''' - {9d_i+1\over 4(d_i+1)} (\tau'')^2 \right)
      +\, \ldots \right] .\qquad
\eea
Then using $\sqrt{\gamma}=e^{f/2},\ \phi=1/r^{d_i}$ and
(\ref{confAdS2-bc}), the Gibbons-Hawking term becomes
\be\label{Ksoft3}
S_{GH} = - {1\over 8\pi G_2} \int d\tau\,\sqrt{\gamma}\,\phi\, K =
- {1\over 8\pi G_2} \int {d\tau\over r^{(3d_i+1)/2}}\, K
= - {1\over 8\pi G_2}
\int {d\tau \over \epsilon^{d_i+1}\,(\tau')^2}\, \Big[ \ldots \Big]
\ee
where $[\ldots]$ here is the term in square brackets in (\ref{Ksoft2}).
For $d_i=1$ we see that the second term in (\ref{Ksoft2}) is the
Schwarzian known in $nAdS_2$.
More generally we see that the second term ($O(\epsilon^2)$) in
(\ref{Ksoft2}) gives a ${1\over\epsilon^{d_i-1}}$ divergence after
sticking it into (\ref{Ksoft3}).\ Using\ $\nabla^2\phi =
{1\over\sqrt{\gamma}}\del_\tau(\sqrt{\gamma}\gamma^{\tau\tau}\del_\tau\phi)
= e^{-f/2}\del_\tau(e^{-f/2}\del_\tau\phi)$\ as well as
$\del_\tau={1\over\tau'}\del_u$ and (\ref{confAdS2-bc}), we note that 
this can be cancelled by a countertem of the form
\be
\int d\tau\sqrt{\gamma}\,\phi^{-{d_i+1\over 2d_i}}
   \left( - (\nabla\phi)^2 - \phi\nabla^2\phi \right)\
   \propto\ \int {d\tau\over\epsilon^{d_i-1}}\, (\tau')^{{-4d_i\over d_i+1}}
   \left( {\tau'''\over\tau'} - {5d_i+1\over 2(d_i+1)}\,
        {(\tau'')^2\over(\tau')^2}\, \right)
\ee
in the spirit of holographic renormalization. There is structural
similarity with\ $\int {\cal R}$ in higher dimensions and its reduction
(\ref{curv-redux}).
For general $d_i$, there are further subleading divergences. Focussing
on $d_i=3$ which is the $AdS_5$ reduction, we see that the $O(\epsilon^4)$
term in the expansion (\ref{Ksoft2}) cancels the overall
${1\over\epsilon^4}$ divergence giving a finite term at this order:
this is the term in the second line in (\ref{Ksoft2}).
Note that this
$O(\epsilon^4)$ subleading term is at the same order as the
term in (\ref{Tttren-AdSDbb}) which gives the black hole excitation:
in some ways this is in sync with the expectation of soft modes
arising from the reduction of low lying hydrodynamic modes in the
Super Yang-Mills CFT dual to the $AdS_5$ black brane upstairs. It
would be interesting to understand this systematically.


\section{An extra scalar $\Psi$}\label{sec:Psi}

Now we add an extra scalar field $\Psi$ to (\ref{2dg-gen}): one way
to obtain this consistently is by reduction of the higher dimensional
action\ $S = {1\over 16 \pi G_D} \int d^Dx \sqrt{-g^{(D)}}\,
\big( {\cal R} - {1\over 2} (\del\Psi)^2 - V \big)$, which arises \eg\
for nonconformal branes, $\Psi$ encoding the running gauge coupling.
As before, the potential $V(g,\Psi)$ also contains metric data.
The reduction (\ref{redux+Weyl}) gives
\cite{Kolekar:2018sba,Kolekar:2018chf,Bhattacharya:2020qil}
\be\label{actionXPsiU}
S= {1\over 16\pi G_2} \int d^2x\sqrt{-g}\, \Big(\phi\mathcal{R}
- U(\phi,\Psi) -\frac{1}{2} \phi (\partial\Psi)^2 \Big)\ ,
\ee
with $U=V\phi^{1/d_i}$. The Weyl factors cancel in\
$(\del\Psi)^2=g^{\mu\nu}\del_\mu\Psi\del_\nu\Psi$\ with $\sqrt{-g}$.\
The dilaton potential $U(\phi,\Psi)$ now possibly couples the dilaton
$\phi$ to $\Psi$.\ The equations of motion are
\bea\label{2dimseom-EMD0-Psi}
g_{\mu\nu}\nabla^2\phi-\nabla_{\mu}\nabla_{\nu}\phi
  +\frac{g_{\mu\nu}}{2}\Big(\frac{\phi}{2}(\partial\Psi)^2+U\Big)
  -\frac{\phi}{2}\partial_{\mu}\Psi\partial_{\nu}\Psi = 0\ ,\qquad\quad &&
  \nonumber\\ [1mm]
\mathcal{R}-\frac{\partial U}{\partial\phi}
  -\frac{1}{2}(\partial\Psi)^2 = 0\ ,\qquad\qquad
\frac{1}{\sqrt{-g}}\partial_{\mu}(\sqrt{-g}\,\phi \partial^{\mu}\Psi)
  -\frac{\partial U}{\partial\Psi} = 0\ ,&&
\eea
giving
\bea\label{2dimseom-EMD1-Psi}
(tr)&& \qquad  \del_t\del_r\phi 
- {1\over 2} f'\del_t\phi - {1\over 2} {\dot f} \del_r\phi
+ {\phi\over 2}{\dot\Psi} \Psi' = 0\ , \nn\\
(rr+tt) && \ \ \
- \del_t^2\phi - \del_r^2\phi + {\dot f} \del_t\phi + f' \del_r\phi
- {\phi\over 2} ({\dot\Psi})^2 - {\phi\over 2} (\Psi')^2 = 0 ,\qquad \nn\\
[1mm]
(rr-tt) && \ \ \
- \del_t^2\phi + \del_r^2\phi + e^f U = 0\ ,\\ [1mm]
(\phi)&& \quad\  \big( {\ddot f} - f'' \big)
- {1\over 2} (-({\dot\Psi})^2+(\Psi')^2)
- e^f \frac{\partial U}{\partial\phi} = 0 ,\nn \\
(\Psi)&& \quad\   - \del_{t}(\phi\del_t\Psi) + \del_{r}(\phi\del_r\Psi)
- e^f {\del U\over\del\Psi} = 0\ , \nn 
\eea
in conformal gauge $g_{\mu\nu}=e^f\eta_{\mu\nu}$.
There is nontrivial dynamics in the theory (\ref{actionXPsiU}) driven
by the extra scalar $\Psi$\ \cite{Bhattacharya:2020qil}: in particular
these equations now admit nontrivial cosmological singularities which
can be thought of as the reduction of higher dimensional Big-Crunch
singularities \cite{Das:2006dz}-\cite{Awad:2008jf}. They were analysed
in \cite{Bhattacharya:2020qil} by using power-law scaling ansatze
similar to (\ref{X^2e^f-gen}), but with\
$\phi=t^kr^m,\ e^f=t^ar^b,\ e^\Psi=t^\al r^\beta$,\ in the
time-dependent case. It was shown there that the near singularity
behaviour is universal, giving $k=1,\ a={\al^2\over 2}$\,, the dilaton
potential becoming irrelevant.

For now we will review just time-independent backgrounds: consider
the potential
\be
U(\phi,\Psi) = A \phi^n e^{B\Psi}\, \qquad A<0,\quad B\geq 0\ ,
\ee
in the 2-dim theory (\ref{actionXPsiU}). For $B=0$ and $A=2\Lambda$,
this is the same as $U$ in (\ref{X^2e^f-gen}) in the theory
(\ref{2dg-gen}). For $B\neq 0$, the form of the potential arises in
\eg\ nonconformal branes and more general nonrelativistic theories.
With a power-law scaling ansatz, the equations (\ref{2dimseom-EMD1-Psi})
then give 
\bea
\phi=r^m,\ e^f=r^b,\ e^\Psi=r^\beta\ \quad\Rightarrow\quad\ \
-m(m-1)+bm-{\beta^2\over 2} = 0 ,\quad\ \ m(m-1) + A = 0 , \nn\\
b + m (n - 1) + B\beta + 2 = 0 , \quad\ \
b-{\beta^2\over 2} - An = 0 ,\quad\ \ \beta (m-1) - AB = 0 .\quad
\eea
For $B=0$, these are the same as (\ref{X^2e^f-gen}): we mention that
there are nontrivial time-dependent solutions with $B=0$\
\cite{Bhattacharya:2020qil}. For $B\neq 0$, we obtain
\bea
-m(m-1)+bm-{\beta^2\over 2} = 0 ,\qquad \beta+Bm=0 ,\qquad
b-{\beta^2\over 2} + m(m-1) n = 0\ ,\nn\\
\ra\ b=m(n+1)\,,\ \ \ \beta = \sqrt{2m(mn+1)}\,,
\ \ \ B=\sqrt{{2(mn+1)\over m}}\quad\Rightarrow\quad
m = {2\over B^2 - 2n}\ .\ \
\eea
We know that\ $\phi$ grows as $r$ decreases so $m<0$: this is
consistent with $m(m-1)=-A>0$ and gives nontrivial consistency
conditions for the existence of these backgrounds
\be\label{condns-BA}
B^2<2n\ ,\qquad\quad
{2\over B^2 - 2n} \left( {2\over B^2 - 2n} - 1 \right) = -A > 0\ .
\ee
(The first condition is also consistent with null energy conditions.)
Our discussion here is bottom-up. Of course these conditions are all
satisfied in the reduction of known higher dimensional theories as
noted in \cite{Bhattacharya:2020qil}: \eg\ we have\ $n={1\over d_i}$ and
$B=\sqrt{{2(-\theta)\over d_i(d_i-\theta)}}$\ with $\theta\leq 0$,\
and\ $m=-(d_i-\theta),\ \ b=-(d_i-\theta)(1+{1\over d_i})$\,; also\
$A=(d_i-\theta)(d_i+1-\theta)$. The higher dimensional space is then
(\ref{hvLif}). As stated there, some of the known reductions include
those of nonconformal $Dp$-branes: the 2-dim geometries regulated with
a blackening factor are then the reductions of black nonconformal
$Dp$-branes \cite{Itzhaki:1998dd}.

Now we make a few comments on the energy-momentum tensor in
(\ref{actionXPsiU}). From the 2-dim equations of motion
(\ref{2dimseom-EMD0-Psi}), we have\ ${\cal R}=\del_\phi U+{1\over
  2}(\del\Psi)^2$. Thus the $(\del\Psi)^2$ terms cancel on-shell: if
the potential $U(\phi)$ does not contain $\Psi$, then the scalar has
entirely disappeared from the on-shell action, which is in fact of the
same form as (\ref{Sren}): in particular the same counterterm there
suffices. We will not discuss more complicated dilaton potentials
$U(\phi,\Psi)$ here.

Now we consider the 2-dim cosmology obtained from the $AdS_D$ Kasner
reduction in \cite{Bhattacharya:2020qil}:
\bea\label{AdSDK-2d}
&& \qquad ds^2 = {1\over r^2} (-dt^2 + dr^2) + {t^{2p_i}\over r^2} dx_i^2\,,
\qquad e^\Psi = t^\al \nn\\
\ra\quad && \phi={t\over r^{d_i}}\,,\qquad
ds^2={t^{(d_i-1)/d_i}\over r^{d_i+1}}(-dt^2+dr^2)\,,\qquad
  e^\Psi=t^{\sqrt{2(d_i-1)/d_i}}\ .\quad
\eea
Noting that the extra scalar required to drive these cosmologies has
disappeared in the on-shell action, we will see from (\ref{Tttren})
that the energy-momentum tensor $T^{ren}_{tt}$ in vanishes. The
scaling ansatz $\phi=t^kr^m,\ e^f=t^ar^b,\ e^\Psi=t^\al r^\beta$,
present in these cosmological backgrounds shows that the
$t$-dependence in (\ref{AdSDK-2d}) appears solely in terms of
multiplicative factors over the $AdS_D$ background profile
(\ref{AdSD-2dmetDil}) itself.  Thus the holographic energy-momentum
tensor (\ref{Tttren}) is
\bea
&& ds^2=e^f(-dt^2+dr^2)\,,\qquad e^f={t^a\over r^{d_i+1}}\,,\quad
\phi={t^k\over r^{d_i}}\,;\qquad a={d_i-1\over d_i}\,,\quad k=1\ ,\nn\\
&& T^{ren}_{tt} = {\epsilon^{{d_i+1\over 2}}\over 8\pi G_2}
\left(-\epsilon^{{d_i+1\over 2}}\,{(-d_i)\over\epsilon^{d_i+1}}\,(t^{-{a\over 2}+k})
- {d_i\over\epsilon^{{d_i+1\over 2}}}\, (t^{{k(d_i+1)\over 2d_i}}) \right)
{-1\over\epsilon^{d_i+1}} = 0\ .
\eea
The $t$-dependences in both factors are in fact identical since the
algebraic conditions on the exponents following from the equations of
motion \cite{Bhattacharya:2020qil} give\ $a=k({d_i-1\over d_i})$\ which
is seen to be satisfied above. Thus this of the same form as
(\ref{Tttren-AdSD}) apart from identical $t$-factors in each term
and leads to $T^{ren}_{tt}=0$.
This is also corroborated by (\ref{Ttt-D}) in the higher dimensional
$AdS_D$ Kasner spacetime \cite{Das:2006dz}-\cite{Awad:2008jf}: we see
that the t-dependence disappears in $\phi$-terms and $g^{rr}$ has
none. Further the other counterterms discussed in \cite{Awad:2007fj}
also cancel in this holographic screen so finally $T_{tt}^{(D)}=0$.
Note that each $t$-factor has a positive exponent so in the vicinity
of the singularity at $t=0$ each term independently becomes
vanishingly small. Overall this suggests a kind of fine-tuning to the
energy-momentum tensor in these holographic screens: although the bulk
fields (in particular the metric) have nonvanishing non-normalizable
deformations turned on, the response vanishes suggesting that the
state is non-generic. However this is screen-dependent: a
Penrose-Brown-Henneaux transformation recasting the metric
schematically as\ $ds^2 = {t^a\over r^{d_i+1}} (-dt^2+dr^2) \equiv
dR^2 ({1\over R^{d_i+1}}+\ldots) + \gamma_{TT}(T,R)dT^2$\ appears
to lead to a nonvanishing energy-momentum tensor. A more detailed
study of this would presumably yield analogs of the corresponding
findings in \cite{Awad:2007fj}.

\subsection{Some comments on scalar probes}\label{sec:scalarProbes}

Now we will discuss aspects of massless scalars regarded as probes:
we will focus on the theory (\ref{2dg-AdSD}) obtained from $AdS_D$
reduction, so as to facilitate direct comparison of the 2-dim and
higher dimensional discussions. To generalize to (\ref{2dg-gen})
with a general monomial potential, it is adequate to set
$d_i\ra {1\over n}$ in most expressions.\\
A massless scalar probe $\psi$ descending from higher dimensions has
action and equation of motion
\be\label{actEOM-psi}
-\int d^2x \sqrt{g}\, \phi_B (\del\psi)^2 \quad\ra\quad
    {1\over\sqrt{g}}
    \del_\mu\big(\phi_B \sqrt{g}\,g^{\mu\nu} \del_\nu\psi\big)=0\ ,
\ee
focussing on the Euclidean theory: here the background dilaton profile
$\phi_B$ is non-constant. For a time-independent probe in the background
(\ref{AdSD-2dmetDil}), we obtain
\be
\del_r(\phi_b\del_r\psi)=0=\del_r(r^{-d_i}\del_r\psi)=0 \quad\ra\quad
\del_r\psi = {C_1\over\phi_B}\quad\ra\quad \psi=C_0+C_1r^{d_i+1}\ .
\ee
So for $d_i>1$, the scalar probe dies rapidly towards the boundary.
The scalar probe perturbation modifies the background potential
which can be read off from the effective action as
\be
\int \left(\phi{\cal R} - U(\phi) - {1\over 2}\phi g^{rr}(\del_r\psi)^2\right)
\quad\Rightarrow\quad
U(\phi_B) \ra\ U(\phi_B) - {C_1^2e^{-f_B}\over 2\,\phi_B}\ .
\ee
This can now be used to understand backreaction: defining\
$\phi=\phi_B+\varphi$\,, gives from (\ref{2dimseom-EMD1-Psi})
\be\label{backeact}
\del_r^2\varphi = -e^{f_B}\,\delta U = {C_1^2\over 2\,\phi_B}\qquad\ra\qquad
\varphi \sim C_1^2 r^{d_i+2}\ ,
\ee
which as $r\ra 0$ is a small correction to the background dilaton\
$\phi_B\sim {1\over r^{d_i}}$ so that the asymptotic background
(\ref{AdSD-2dmetDil}) does not suffer large backreaction. Equivalently
the background can support small excitations.

Now consider operators dual to scalar probes in the Euclidean
space (\ref{AdSD-2dmetDil}): (\ref{actEOM-psi}) gives
\be
0=\del_\mu(\phi \sqrt{g}\,g^{\mu\nu} \del_\nu\psi)
=\del_\tau(\phi\del_\tau\psi)+\del_r(\phi\del_r\psi) \ \ \
\longrightarrow\ \ \
\del_r({1\over r^{d_i}}\del_r{\psi_\omega})
-{1\over r^{d_i}}\omega^2{\psi_\omega}=0\ ,
\ee
with\ $\psi(\tau,r)=e^{-i\omega\tau}\psi_\omega(r)$\,.
This has many similarities with the higher dimensional $AdS_D$
but it is instructive to look at the simplest nontrivial case $d_i=2$
in detail: then\ ($\nu={d_i+1\over 2}$)
\be
\psi_\omega=\psi_\omega^0\,{r^{(d_i+1)/2}\,K_\nu(\omega r)\over
  \epsilon^{(d_i+1)/2}\,K_\nu(\omega \epsilon)}
\quad\xrightarrow{\ d_i\ra 2\ }\quad
\psi_\omega^0\, {(1+\omega r)\,e^{-\omega r}\over
  (1+\omega\epsilon)\,e^{-\omega\epsilon}}\ ,
\ee
which is regular at large $r$ although the 2-dim space is singular.
Using the bulk equation, the on-shell action gives a boundary term as
usual: expanding in $\epsilon$ gives
\be\label{2ptcorrfn1}
-S \sim \int d\omega\, {1\over \epsilon^{d_i}}\,
\psi_{-\omega}\del_r\psi_\omega\Big|_\epsilon\ \ \ra\ \ 
\int {d\omega\over\epsilon^2}\, \psi_{-\omega}^0 \psi_\omega^0\,
{-\omega^2\epsilon\over (1+\omega\epsilon)}\,
=\, \int d\omega\, \psi_{-\omega}^0 \psi_\omega^0\,
\left( {-\omega^2\over\epsilon}+\omega^3 + \ldots \right)
\ee
The first term can be removed by a local counterterm\
$\int d\tau\sqrt{\gamma}\,\phi^{5/4} (\del\Psi)_B^2$: this in fact arises
from the higher dimensional\
$\int d\tau d^{d_i}x\sqrt{h} h^{ij}\del_i\Psi\del_j\Psi
\ra \int dt\sqrt{\gamma^{(2)}}\,\phi\,\gamma^{(2)\tau\tau} (\del_\tau\Psi)^2$\
after incorporating the Weyl transformation (\ref{redux+Weyl}).
The second nonlocal term is the same momentum space 2-point function
as in $AdS_4$. Fourier transforming, now in $0+1$-dimensions, gives
in position space
\be\label{2ptcorrfn2}
\lan O(\tau) O(\tau')\ran \sim \int d\omega\, e^{i\omega\Delta\tau}\, \omega^3\
\sim\   {1\over \left(\Delta\tau\right)^4}\ ,
\ee
giving the operator dimension\ $\Delta=2$ for $O(\tau)$.
For $d_i>1$, the short-time divergence here ($\Delta\tau\ra 0$) is more
severe than for a dimension $\Delta=1$ operator as would be the case
if the dilaton factor were absent in (\ref{actEOM-psi}). This can be
extended to $d_i>2$ also: the momentum space answer is identical to
the corresponding calculation in higher dimensional $AdS_D$ giving
$\omega^{2\nu}(\log\omega\epsilon)$\ (including the possible
logarithmic piece for $\nu\in\BZ$) but the 1-dim Fourier transform
gives the position space correlator\ ${1\over(\Delta\tau)^{d_i+2}}$ and
thereby $1+{d_i\over 2}$ for the dimension of the operator $O(\tau)$,
which is distinct from $AdS_{d_i+2}$ where the operator dimension
would have been $d_i+1$, from the $(d_i+1)$-dim Fourier transform\
(see also \cite{Taylor:2017dly}).

It is interesting that the above 2-point correlation function calculation
is insensitive to the singularity in the zero temperature 2-dim space,
stemming from regularity of the bulk mode functions. 
It is likely that this is a generic feature, \ie\ low point correlators
do not see the singularity in the zero temperature 2-dim space. It
would be interesting to understand this better.

It would seem overall that we have just scratched the surface here. In
light of (\ref{backeact}) one might think backreaction effects are
small: it would be interesting to study correlation functions in
greater detail, incorporating possible time reparametrizations along
the lines of \cite{Almheiri:2014cka}. Modes with nontrivial dilaton
couplings may also be of interest in understanding aspects of the SYK
theory: see \eg\ the recent work \cite{Das:2020kmt}.

\section{Discussion}\label{sec:Disc}

We have discussed aspects of generic 2-dimensional dilaton gravity
theories bottom-up. The 2-dim geometry is in general conformal to
$AdS_2$ and has IR curvature singularities at zero temperature, which
are regulated by a black hole.  As we saw, the on-shell action is
divergent: we discussed the holographic energy-momentum tensor by
adding appropriate counterterms.  For theories obtained by dimensional
reduction of the gravitational sector of higher dimensional theories,
the 2-dimensional description dovetails with the higher dimensional
one, as we saw explicitly in the reduction of higher dimensional $AdS$
gravity as a concrete example. The analysis of low-lying (``soft'')
modes reveals distinct departures from the $nAdS_2$ Schwarzian,
stemming from further divergences here (\ref{Ksoft2}).  Overall, such
2-dim theories appear to encode higher dimensional gravity
intrinsically (there are parallels with old work
\eg\ \cite{Giddings:1992ae}, also reviews
\eg\ \cite{Strominger:1994tn}). This is quite different from near
extremal objects with $AdS_2\times X$ throats: here the
$X$-compactification leads to JT gravity which is intrinsically
2-dimensional (and topological in some sense).  We also discussed
adding an extra scalar field which drives nontrivial dynamics, in
particular the 2-dimensional cosmological singularities discussed in
\cite{Bhattacharya:2020qil}: the new dynamical direction that the
scalar defines may have similarities with that in backgrounds that
include rotation, gauge fields etc
\eg\ \cite{Castro:2018ffi,Moitra:2018jqs}. For scalar probes with
dilaton couplings, we saw that correlation functions of dual operators
(\ref{2ptcorrfn1}), (\ref{2ptcorrfn2}), are insensitive to the
singularity in the zero temperature 2-dim space, a feature likely to
extend to more general low point correlators.

We now recall the discussion in \cite{Saad:2019lba} of JT gravity as
dual to a random matrix ensemble: further developments appear in
\cite{Stanford:2019vob}-\cite{Alishahiha:2020jko}. In this light, the
perturbative studies here suggest that generic 2-dim dilaton gravity
theories (\ref{2dg-gen}) are somewhat different from JT gravity and
theories ``near JT'', \eg\ through the behaviour of IR singularities
in the zero temperature 2-dim geometry and the divergence in the
on-shell action.  Nonperturbatively in JT gravity, we recall that the
$\phi$ path integral leads to a sum over constant curvature surfaces
of various topologies (which maps to a corresponding expansion of a
matrix integral).
By comparison for a more general potential in the Euclidean theory, we
have\ $\int Dg\,D\phi\; {\rm exp}[\int \sqrt{g}\,(\phi{\cal R} - U)]$
which using the the dilaton equation in (\ref{2dimseom-EMD0}) and
noting (\ref{onshellActionD}) naively leads to\ $\sim \int Dg\, {\rm
  exp}[\#\int\sqrt{g}\, (-{\cal R})^{-1/(d_i-1)}]$\ in the specific
case (\ref{2dg-AdSD}), with $\#$ a positive number. By comparison, a
simple gaussian potential $U=\lambda\phi^2$ gives\ $\sim \int Dg\,
{\rm exp}[\#\int\sqrt{g}\, {{\cal R}^2\over\lambda}]$\ after doing the
gaussian $\phi$-integral, which is 2-dim ${\cal R}^2$ gravity.  These
structures appear more intricate (also, the sign of $U$ enters
nontrivially): it would be interesting to understand them in greater
depth, perhaps towards mapping to some effective matrix model.

It would appear that 2-dim theories such as (\ref{2dg-gen}) are not
``near JT'' in the sense of the deformations of JT gravity in
\cite{Witten:2020ert,Witten:2020wvy}: roughly $|U-U_{JT}|$ here is not
sufficiently small asymptotically, unlike there.  As a concrete
example, the theory (\ref{2dg-AdSD}) for $d_i>1$ in some essential
sense faithfully reflects its higher dimensional origins. Ordinarily
we regard gravity in dimensions $D\geq 4$ as UV completed by a
string/M theory: a particular gauge/gravity dual pair is usually
pinned down by specifying the precise matter content, including
information on $X$ (so the bulk string theory is then critical).  From
this point of view, in treating the $AdS_D$ gravity reduction
(\ref{2dg-AdSD}) as a standalone theory, we have effectively performed
some sort of average or partial trace (loosely speaking) over the
UV-information that encodes the identity of the specific gauge/gravity
dual pair that it was compactified from\ (equivalently, with some
partial trace over local operators from the dual field theory point of
view). In this light, it would perhaps be interesting to understand
minimal string theories which are 2-dimensional to begin with and
their dual matrix models (see \eg\ \cite{Douglas:2003up}, as well as
older reviews \eg\ \cite{Klebanov:1991qa,Ginsparg:1993is})
and the possibly distinct kinds of embeddings of such 2-dim gravity
theories obtained thereby (see
\eg\ \cite{Saad:2019lba,Stanford:2019vob,Johnson:2020exp} for some
discussions, as well as \eg\ \cite{Betzios:2020nry}).  Since the
generic theory (\ref{2dg-gen}) is structurally similar to
(\ref{2dg-AdSD}), this suggests that the generic 2-dim theory
(\ref{2dg-gen}) is a UV-incomplete low energy effective theory, akin
to a thermodynamic ensemble in a sense\ (naively the dual operators
are simply $T_{tt}^{(1)}$ and $T_\phi\equiv T_{ii}$; in the case of a
higher dimensional CFT compactified spatially, these are subject to
the tracelessness condition, thus giving $T_{tt}^{(1)}$ alone in the
effective 1-dim dual, with no microstate information).
While this is adequate for
understanding various holographic (large $N$) observables as we
have seen, the nature of such a thermodynamic ensemble appears
fundamentally different from the ensemble dual to JT. The 2-dim
theories here represent some sort of ``effective holography'',\
\ie\ some approximate gauge/gravity duality, with a subset of
observables: it would be interesting to understand more systematically
such incomplete holographic models, as well as their validity and
usefulness. There are some
parallels with the discussions in \cite{McNamara:2020uza} of higher
dimensional gravity. The case $d_i=1$ (\ref{2dg-AdSD}) is the $AdS_3$
reduction, which as we have seen has many parallels with JT gravity:
this appears in accord with \cite{Maxfield:2020ale}.  It would be
interesting to understand these issues better and gain more insight
into quantifying the ensemble nature of such effective generic 2-dim
gravity theories (at least those that are adequately well-defined
intrinsically or via reductions from reasonable theories upstairs, and
not in some sort of swampland) and possible effective matrix models,
and the role of replica wormholes
\cite{Penington:2019kki,Almheiri:2019qdq} here.

\vspace{7mm}

{\footnotesize \noindent {\bf Acknowledgements:}\ \ It is a pleasure
  to thank Sumit Das, Tom Hartman, Dileep Jatkar, Igor Klebanov, Kedar
  Kolekar, Arnab Kundu, Alok Laddha, Henry Maxfield and Douglas
  Stanford for some early correspondence around Strings, helpful
  discussions and comments on a draft, and the organizers of the
  virtual Strings 2020 for a stimulating conference. This work is
  partially supported by a grant to CMI from the Infosys Foundation.
}

\vspace{-1mm}

\appendix

\section{Reduction details}\label{sec:AppA}

\noindent {\bf $T^{D-2}$ reduction:}\ \ The $D$-dimensional space with
$D=d_i+2$ is of the form (\ref{redux+Weyl}): we have
\be
\int {d^Dx\over G_D} \sqrt{-g^{(D)}}\, ({\cal R}^{(D)}-V)\ \
\ra\ \ \int {d^2x\over G_2} \sqrt{-g^{(2)}}
\Big(\phi {\cal R}^{(2)} + {D-3\over D-2}\, {(\del\phi)^2\over\phi}
- V \phi \Big)\ ,
\ee
from reduction to the 2-dim space $g^{(2)}$.
A total derivative from the reduction cancels one from the reduction of the
Gibbons-Hawking term. As in \cite{Almheiri:2014cka} and
\cite{Kolekar:2018sba,Kolekar:2018chf}, noting
\be
\sqrt{-g^{(2)}} \Big[\phi {\cal R}^{(2)} +
  \lambda {(\del\phi)^2\over\phi}\Big]\ \ \
\xrightarrow{\ g_{ab}^{(2)} = \phi^{-\al}\, g_{ab}\ }\ \ \
\sqrt{-g} \Big[\phi {\cal R} + (\lambda-\al) {(\del\phi)^2\over\phi} \Big] ,
\ee
we use the Weyl transformation in (\ref{redux+Weyl}) to absorb the
dilaton kinetic term into ${\cal R}$, giving (\ref{2dg-AdSD}); the
factor in $V$ gives $\phi^{-{D-3\over D-2}+1}$\,.

\noindent {\bf $S^{D-2}$ reduction:}\ \ Here the curvature of the sphere
gives additional terms (see \eg\ \cite{Strominger:1994tn,Nojiri:2000ja,
  Grumiller:2001ea,Grumiller:2002nm}): we have
\be\label{curv-redux}
R^{(D)} = R^{(2)} + {(D-3)(D-2)\,\phi^{{D-4\over D-2}}\over\phi}
+ {D-3\over D-2}\, {\big(\del_{(2)}\phi\big)^2\over\phi^2} 
- 2 {\nabla_{(2)}^2 \phi\over\phi}\,,
\ee
from (\ref{redux+Weyl}) with $d\sigma_{D-2}^2\equiv d\Omega_{D-2}^2$.
Then reduction and the Weyl transformation (\ref{redux+Weyl}) gives
\bea
\int d^2x\sqrt{-g^{(2)}} \Big(\phi {\cal R}^{(2)} +
     {D-3\over D-2}\, {(\del\phi)^2\over\phi} +
     (D-2)(D-3) \phi^{{D-4\over D-2}} - V\phi \Big) \nn\\
     \xrightarrow{\ V=2\Lambda\ }\ \
     \int d^2x\sqrt{-g} \Big(\phi {\cal R} + d_i(d_i+1) \phi^{1/d_i}
     + d_i(d_i-1) \phi^{-1/d_i} \Big)\ .
\eea
For global $AdS_D$, the reduction (\ref{redux+Weyl}) of the background gives
\be
ds^2=-(1+r^2)dt^2+{dr^2\over 1+r^2}+r^2d\Omega_i^2 \ \ra\
\phi=r^{d_i}\,,\ \ e^f=r^{d_i-1} \Big(-(1+r^2)dt^2+{dr^2\over 1+r^2}\Big) ,
\ee
with curvature\ ${\cal R}=-r^{-d_i-1} ((d_i+1)r^2-(d_i-1))$\,. Thus 
${\cal R}\sim {1\over r^{d_i-1}}$ as $r\ra 0$ (interior) so the
singularity structure is similar to that for the $AdS_D$ Poincare
reduction. The $AdS$ Schwarzschild black hole under the reduction
(\ref{redux+Weyl}) gives
\be
\phi=r^{d_i}\,,\quad 
e^f=r^{d_i-1} \Big(-(1+r^2-{1\over r^{d_i-1}})dt^2
+{dr^2\over 1+r^2-{1\over r^{d_i-1}}}\Big)\ .
\ee

\footnotesize{

}

\end{document}